\documentclass[10pt]{article}
\usepackage{amsmath}
\usepackage{amssymb}
\usepackage{graphicx}

\def\aa{A\&A }

\def\aj{AJ }

\def\ajs{AJSupp}
\def\apj{ApJ}

\begin{document}

\setcounter{figure}{0}
\setcounter{table}{0}
\setcounter{footnote}{0}
\setcounter{equation}{0}

\vspace*{0.5cm}

\noindent {\Large THE SECULAR ABERRATION DRIFT AND FUTURE CHALLENGES FOR VLBI ASTROMETRY}
\vspace*{0.7cm}

\noindent\hspace*{1.5cm} O.\ TITOV$^1$ \\
\noindent\hspace*{1.5cm} $^1$ Geoscience Australia \\
\noindent\hspace*{1.5cm} PO Box 378, Canberra, 2601, ACT, Australia\\
\noindent\hspace*{1.5cm} e-mail: oleg.titov@ga.gov.au\\

\vspace*{0.5cm}

\begin{abstract}

The centrifugial acceleration of the Solar system, resulting from
the gravitational attraction of the Galaxy centre, causes a phenomenon
known as 'secular aberrration drift'. This acceleration of the Solar system
barycentre has been ignored so far in the standard procedures
for high-precision astrometry. It turns out that the current
definition of the celestial reference frame as epochless and
based on the assumption that quasars have no detectable proper
motions, needs to be revised. In the future, a realization of the celestial reference
system (realized either with VLBI, or GAIA) should correct source coordinates from
this effect, possibly by providing source positions together with their proper motions.
Alternatively, the galactocentric acceleration may be incorporated into the
conventional group delay model applied for VLBI data analysis.

\end{abstract}

\vspace*{1cm}


\section{Secular Aberration Drift}


\subsection{Introduction}


This barycentre reference system was adopted by the International Astronomical Union as
the International Reference System (ICRS). The Very Long
Baseline Interferometry (VLBI) technique measures precise group
delays difference in arrival times of radio waves at two radio
telescopes and produces very accurate coordinates of the reference radio sources. The ICRS
is based on a set of theoretical concepts. Practical realization of the ICRS - International
Celestial Reference Frame (ICRF) is presented in a form of astrometric catalogue
of the reference radio source accurate coordinates.

The second realization of the ICRF (ICRF2) contains 295 ``defining" sources
(Fey, Gordon and Jacobs, 2009). The floor error of the most
observed radio sources among the 295 ICRF2 ``defining" is about 41 $\mu$as. The median
error for 1217 radio sources (295 ``defining" and 922 ``non-defining") was found to
be 174 $\mu$as in right ascension and 194 $\mu$as in declination.

All the reductions for high-precision VLBI data are done in the
reference system with an origin in the barycentre of the Solar
system.  In this definition, the axes of the ICRS are defined by
the adopted positions of a specific set of extragalactic objects
(presumably, very distant quasars), which are assumed to have no
measurable proper motions. However, the acceleration of the Solar
System barycenter would cause a dipole systematic effect in the
proper motion described by the first order electric type vector
spherical harmonic with magnitude of 4--6 $\mu$as/yr. After an
early mention by Fanselow (1983), this effect known as secular
aberration drift (SAD) was discussed in more detail later (e.g.
Bastian, 1995; Gwinn et al., 1997; Sovers et al., 1998; Kovalevsky,
2003; Kopeikin and Makarov, 2006).  Finally, it has been confirmed,
for the first time, by Titov, Lambert \& Gontier (2011) by the
least squared adjustment of the apparent proper motion of 555
reference radio sources. Its magnitude, $6.4\pm1.5~\mu$as/yr,
corresponds to the Galactocentric acceleration of
$(3.2\pm0.7)\times10^{-13}$~km/s$^2$, in good agreement with the
theoretical predictions.

\begin{figure}[ht]
\begin{center}
\includegraphics[scale=0.5]{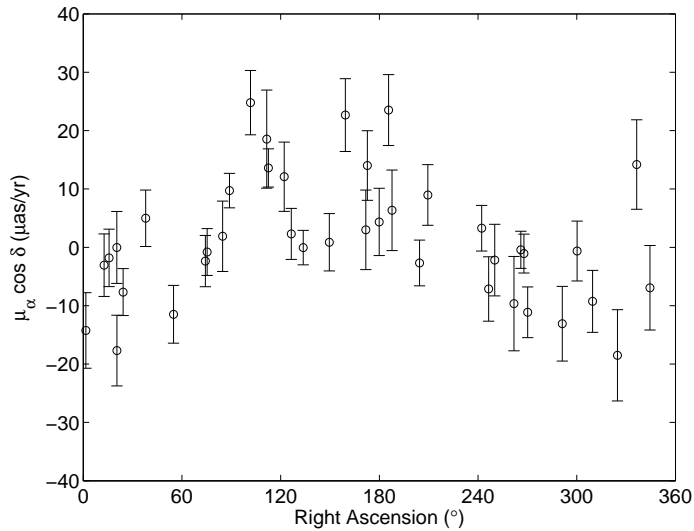}
\caption{The proper motion in right ascension vs.\ the right ascension of 40 sources observed in more than 1000 sessions.\label{Titov1_01}}
\end{center}
\end{figure}

\begin{figure}[ht]
\begin{center}
\includegraphics[scale=0.5]{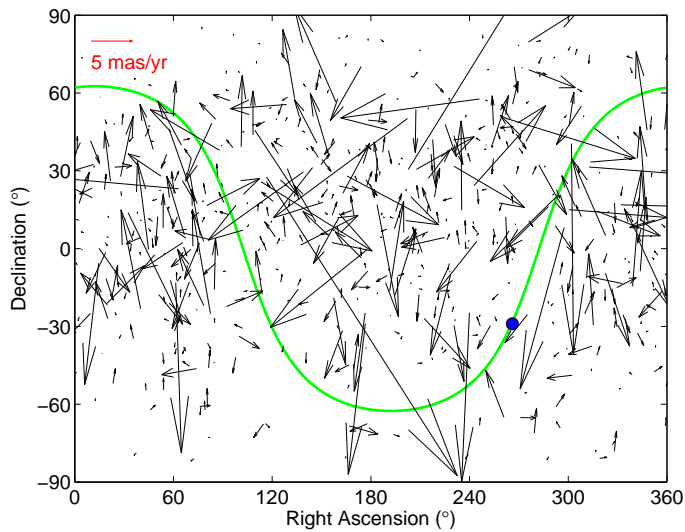}
\caption{The proper motions of the 555 sources. The green curved line
  represents the equator of the Milky Way, whose center is indicated
  by the large blue point. \label{Titov1_02}}
\end{center}
\end{figure}

\begin{figure}[ht]

\begin{center}
\includegraphics[scale=0.5]{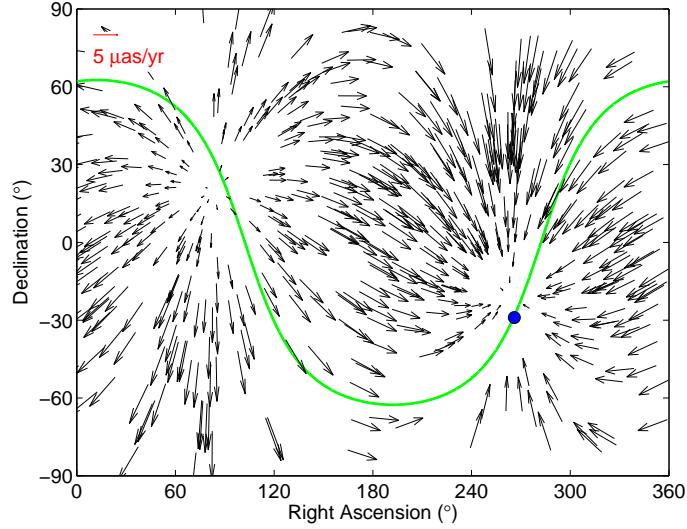}
\caption{The estimated dipole component of the velocity field.\label{Titov1_03}}
\end{center}
\end{figure}

Figure~\ref{Titov1_01} displays $\mu_{\alpha}\cos\delta$ versus $\alpha$ for the
40 astrometricaly stable radio sources observed in more than
1,000 sessions.  The plot reveals dipole systematic
effect even without adjustment by the least squares.
Individual proper motions of 555 radio sources are shown in
Figure~\ref{Titov1_02}.  Though, no systematic effect is
observed in this plot, the least squares found the dipole as on
Figure~\ref{Titov1_03}.

Independent analysis of the SAD was done using 328 individual proper
motions obtained by Vladimir Zharov using the alternative
software ARIADNA (Zharov et al. 2010). The magnitude estimate is
$4.8\pm1.4~\mu$as/yr, and coordinates of the acceleration vector are estimated
as $270^{\circ} \pm 30^{\circ}$  in right ascension and $-54^{\circ} \pm 17^{\circ}$  in declination.
The consequences of the secular aberration drift discovery are discussed
in this paper.



\section{Future Challenges For VLBI Astrometry}


In accordance with the IAU Resolutions, the adopted International Celestial Reference System (ICRS)
is quasi-inertial, i.e. its fundamental axes do not rotate. However, acceleration of the origin is
allowed, in opposition to the definition of the inertial
reference system. This theoretical concept of the quasi-inertial
system is realised in a form of the catalogue of accurate
coordinates of extragalactic radio sources, known as ICRF2 (Fey,
Gordon and Jacobs, 2009).  Directions of the fundamental axes
are fixed by positions of the 295 ICRF2 ``defining" radio
sources.  This theoretical concept of quasi-inertiality as well
as practical efforts for production of new catalogues are used
to focus on control of the net rotation which may be
induced by the variations of the intrinsic structure of the
radio sources. All the non-rotational systematic effects in
proper motions were considered to be negligibly small and,
eventually, ignored. As a result, the No-Net-Rotations
constraints which are commonly used to suppress the forbidden
rotational systematic also suppress the dipole systematic
effect, theoretically allowed.  Nonetheless, due to
unprecendented improvement of the geodetic VLBI data precision,
the dipole systematic effect should not be ignored. Otherwise,
the conventional reductional model would become increasingly
inadequate. Therefore, the geodetic VLBI faces three challenges
which should be solved in near future.


\subsection{Challenge 1. How To Implement This Effect?}



The dipole proper motion induced by the SAD would result in displacement of the actual positions
of the reference radio sources from the positions published in catalogues, i.e. ICRF2. This
displacement may reach 130 $\mu$as over the 20-year period of
observations, exceeding the 3-$\sigma$ level of the ICRF2 floor error
(41 $\mu$as).  If this effect is ignored, estimates of some other
parameters may be biased, for instance, corrections to the nutation
angles, or the variations UT1-UTC estimated from the Intensive
sessions.  Therefore, it is reasonable to discuss a possible
to implement the SAD in the conventional procedure of VLBI data
reductions.

One possibility is to introduce the dipole proper motion
for all reference radio sources as an extension to their ICRF2
coordinates. For a distant body of equatorial coordinates 
$(\alpha,\delta)$, the proper motions are estimated as follows

\begin{align} \label{dip1} 
\Delta\mu_{\alpha}\cos\delta &= -d_1\sin\alpha+d_2\cos\alpha, \\
\Delta\mu_{\delta} &= -d_1\cos\alpha\sin\delta-d_2\sin\alpha\sin\delta+d_3\cos\delta, \nonumber
\end{align}

where the $d_i$ are the components of the acceleration vector in unit of the proper motion
calculated as

\begin{align} \label{dip2}
d_1 &= d\cos\alpha_0\cos\delta_0 \\ 
d_2 &= d\sin\alpha_0\cos\delta_0 \nonumber \\ 
d_3 &= d\sin\delta_0, \nonumber
\end{align}

where $d$ is the magnitude of the SAD, and ($\alpha_0$,
$\delta_0$) --- apparent equatorial coordinates of the Galactic
center.

This approach will result to moving away from the attractive
idea of the treatment the extragalactic radio sources as fixed
fiducial points on the celestial sphere. It may not be
appropriate to introduce the systematic proper motion for the
radio reference frame only, but keeping in mind the future space
astrometric missions (i.e. GAIA, see Perryman et al. (2001),
Mignard (2002)), it should be considered as one of the possible
options.

Another possibility is to revise the conventional group delay
model developed about 20 years ago and recommended by the IERS
and IAU (Kopeikin, 1990; Soffel et al. 1991).  It was shown
(Titov 2010) that SAD may be implemented by the replacement of
the barycentric velocity vector $\vec V$ by the sum $\vec V +
\vec a \Delta t$, where $\vec a$ is the vector of Galactocentric
acceleration and $\Delta t$ is the period of time since an
initial epoch.  This approach keeps the extragalactic radio
sources technically free of the apparent trasversal motion.
However, it is suitable only for the reduction of the geodetic
VLBI data, and it is not clear how to proceed with the data from
the space astrometric missions.  More detailed analysis of this
problem should be initiated to determine the most appropriate
solution.



\subsection{Challenge 2. Are There More Systematic Effects?}

Apart from the SAD, more systematic effects may be detected, although with a smaller chance of success.
The quadrupole component may come either from the Hubble constant anisotropy or the primordial
gravitational waves (i.e. Kristian \& Sachs 1966; Eubanks,
1991; Pyne et al., 1996; Gwinn et al.  1997; Book \& Flanagan,
2010).  Observational results impose constraints on the upper
limit of the quadrupole systematic effect (3 $\mu$as/yr for a
3$\sigma$ standard error (Titov, Lambert \& Gontier 2011). The
amplitude spherical harmonics were estimated to be very close to
the 3$\sigma$ level, therefore, it will be useful to check these
parameter after collection of more observational data.

The rotational harmonics are unlikely to be separated from the Earth orientation parameters.
Although, the differential rotation possible within of a scope of the
formula by Kristian and Sachs (1966),

\begin{align} \label{eq3}
\begin{split}
\frac{de^\mu}{dt}  = & h^{\mu\nu} \left \{ \vphantom{\frac{1}{2}} e^{\beta}(\sigma_{\nu\beta}
  + \omega_{\nu\beta}) + r \left [ \vphantom{\frac{1}{2}} e^{\beta}(\sigma_{\nu\beta}
+\omega_{\nu\beta}) u_{\mu\nu} - e^{\beta} E_{\mu\nu} \right. \right.\\
& \left. \left.  + \frac{1}{2} e^\beta e^\gamma (u_{\nu\beta\gamma} -
\epsilon_{\nu\beta\gamma} H_{\mu\nu})+ e^{\beta}e^{\gamma}e^{\lambda}
(\sigma_{\nu\gamma} + \omega_{\nu\gamma})\sigma_{\beta\lambda} \right ]+ ... \right \}
\end{split}
\end{align}

\noindent may be disclosed; where $\sigma$ - notation for the shear
tensor, $\omega$ - for the rotation tensor, $E$, and $H$ - for
the tensors describing the electric and magnetic-type
gravitational waves, respectively.  The distance $r$ before the
squared brackets means that if at least one of the tensors in
not negligible at level of $\mu$as/yr, then the corresponding
component will increase for more distant radio sources. Kristian and Sachs (1966)
considered the case of gravitational waves with wavelengths of the Universe size
$\lambda_{Gr} \sim R_{Un}$. Later papers argued that if $\lambda_{Gr} < R_{Un}$,
then the corresponding spherical harmonics are independent on the
distance (Pyne, 1996; Gwinn et al., 1997; Book \& Flanagan, 2010).
So, the first part of (3) should be developed as follows

\begin{equation} \label{eq4}
\frac{de^\mu}{dt}  =  h^{\mu\nu} \left \{ \vphantom{\frac{1}{2}} e^{\beta}(\sigma_{\nu\beta}
  + \omega_{\nu\beta}) - e^{\beta} E_{\mu\nu}  + \frac{1}{2} e^\beta e^\gamma (u_{\nu\beta\gamma} -
\epsilon_{\nu\beta\gamma} H_{\mu\nu}) \right \}
\end{equation}

The spherical harmonics of order 3 (octopole) and higher
accomodate only 15\% of the total power of the gravitational
waves (Gwinn et al., 1997), and we have not found statistically
significant harmonics in the proper motion analysis.
Nonetheless, these higher harmonics may be still studied in
future. To estimate the distance-dependent effects in
the spherical harmonics, more proper motions at redshift z$>$2
are required to be measured.




\subsection{Challenge 3. ``Observational''}

The proper motions of 555 radio sources have been used so far to
estimate the SAD components by use of the least squares method,
increasing the figure to 2000 will improve the formal accuracy
by a factor of 2. In fact, several thousand radio sources
were observed in 1 or 2 VCS sessions (VLBI Calibrator Survey)
for the last decade (i.e.  Beasley et al., 2002). Therefore, we
are able to calculate new proper motions by undertaking regular
observations of the VCS sources in furter 1 or 2 sessions
separated from the original epochs by a sufficient time span.

The mutual correlation between different spherical harmonics
will vanish for homogeneous coverage of the celestial
sphere with observational data. Unfortumately, the  shortage of
proper motions for the radio sources around the South Pole
leads to a correlation between parameters in the matrix of
normal equations (Titov \& Malkin, 2009).  This correlation
could be reduced by increasing the data at this zone of the
celestial sphere. It is planned to use the Australian VLBI
network (AuScope) and New Zealand station Warkworth (Lovell
et al., 2010) for regular observations of those radio sources which
have comparatively strong flux ($>$0.4 Jy) but are observed
only occassionally.


\section{CONCLUSION}

Appearance of the systematic effects in the apparent proper motion of extragalactic radio sources
observed with Very Long Baseline Interferometry requires a revision of some basic assumptions
used for many years to establish the fundamental reference frame. The
gravitational acceleration of the Solar system barycenter needs to be incorporated to the
analytical models for high-precision astrometric data reduction. More work needs to be done
in theory, in observations and in data analysis to reveal the small signals with confidence.
In particular, observations of the quasars in the southern
hemisphere are highly essential.


\section{ACKNOWLEDGMENT}

We are thankful to Vladimir Zharov for providing the independent
results from the ARIADNA software and Laura
Stanford for help in preparation of the manuscript. This paper
has been published with permisson of the Geoscience Australia
CEO.

\vspace*{0.7cm}

\noindent {\large 5. REFERENCES}
%
%
%
%
%
{

\leftskip=5mm
\parindent=-5mm

\smallskip

Bastian, U. 1995, ``Direct detection of the Sun's galactocentric
acceleration'', In: M. A. C. Perryman \& F. Van Leeuwen (Eds.),
Proc. RGO-ESA Workshop on Future Possibilities for Astrometry in
Space, ESA SP-379, 99. 

Beasley, A. J., Gordon, D., Peck, A. B., et al. 2002, ``The VLBA
Calibrator Survey-VCS1'', \ajs 141, pp. 13--21. 

Book, L.G., Flanagan, E., 2011, ``Astrometric Effects of a Stochastic
Gravitational Wave Background'', Phys. Rev. D, 83(2). 

Eubanks, T. M., 1991, ``A consensus model for relativistic effects in geodetic VLBI'', In: Proceedings of the U.S. Naval
Observatory Workshop on Relativistic Models for use in Space Geodesy,
pp. 60--82. 

Eubanks, T. M., 1991, ``Astrometric Cosmology using Radio
Interferometry'', In: Proc. of 179th AAS Meeting, Atlanta, Georgia,
p. 1465. 

Eubanks, T. M., Matsakis, D. N., Josties, F. J., et al. 1995,
``Secular motions of extragalactic radio-sources and the stability of
the radio reference frame'', In: E. Hog \& P. K. Seidelmann (Eds.),
International Astronomical Union (IAU) Symp. 166, Kluwer Academic,
Publishers, Dordrecht, p. 283. 

Fanselow, J. L. 1983, ``Observation Model and Parameter Partials for
the JPL VLBI Parameter Estimation Software MASTERFIT-V1.0'', JPL
Publication 83-39. 

Fey, A. L., Gordon, D. G., \& Jacobs, C. S. (Eds.) 2009, ``The Second
Realization of the International Celestial Reference Frame by Very
Long Baseline Interferometry'', Presented on behalf of the IERS / IVS
Working Group, International Earth Rotation and Reference Systems
Service (IERS) Technical Note 35, Frankfurt am Main: Verlag des
Bundesamts f\"ur Kartographie und Geod\"asie. 

Gwinn, C. R., Eubanks, T. M., Pyne, T., et al., 1997, ``Quasar proper
motions and low-frequency gravitational waves'', \apj  485,
pp. 87--91. 

Kopeikin, S. M., \& Makarov, V. V. 2006, ``Astrometric effects of
secular aberration'', \aj 131, pp. 1471--1478. 

Kopeikin, S. M., 1990, ``Theory or relativity in observational radio
astronomy'', Sov. Astron. 34(1), pp. 5--10. 

Kovalevsky, J. 2003, ``Aberration in proper motions'', \aa 404,
pp. 743--747. 

Kristian, J., \& Sachs, R. K. 1966, ``Observations in cosmology'',
\apj, 143, pp. 379--399. 

Lovell, J., Dickey, J., Gulyaev, S., et al. 2010, ``The AuScope
project and Trans-Tasman VLBI'', In: D. Behrend \& K. D. Baver (Eds.),
Proc. of the 2010 IVS 6th General Meeting, pp. 50--54. 

MacMillan, D. S. 2005, ``Quasar Apparent Proper Motion Observed by
Geodetic VLBI Networks'', In: J. Romney \& M. Reid (Eds.), Future
Directions in High Resolution Astronomy: The 10th Anniversary of the
VLBA, ASP Conference Proceedings, San Francisco: Astronomical Society of the Pacific, pp. 477--491. 

Mignard, F. 2002, ``Fundamental Physics with GAIA'', In: O. Bienaym\'e
\& C. Turon (Eds.), GAIA: A European Space Project, EAS Publication
Series, 2, pp. 107--121. 

Perryman, M. A. C., de Boer, K. S., Gilmore, G., et al. 2001, ``GAIA:
Composition, formation and evolution of the Galaxy'', \aa  369,
pp. 339--363. 

Pyne T., Gwinn, C.R., Birkinshaw, M., et al., 1996, ``Gravitational
radiation and Very Long Baseline Interferometry'', \apj, 465,
pp. 566--577. 

Soffel, M. H., Muller, J., Wu, X., \& Xu. C., 1991, ``Consistent
Relativistic VLBI Theory with Picosecond Accuracy'', \aj 101,
pp. 2306--2310.  

Sovers, O. J., Fanselow, J. L., Jacobs, C. S. 1998, ``Astrometry and
geodesy with radio interferometry: experiments, models, results'',
Rev. Mod. Phys. 70, pp. 1393--1454. 

Titov, O., 2009, ``Systematic effects in the radio source proper
motion'', In:  G. Bourda et al. (Eds.), Proc. 19th European VLBI for
Geodesy and Astrometry (EVGA) Working Meeting, pp. 14--18. 

Titov, O., 2010, ``Estimation of the Acceleration of the Solar-System
Barycenter Relative to a System of Reference Quasars'', Astronomy
Reports 55, pp. 91--95. 

Titov, O., Malkin, Z., 2009, ``Effect of asymmetry of the radio source
distribution on the apparent proper motion kinematic analysis'', \aa
506, pp. 1477--1485. 

Titov, O., Lambert, S., \& Gontier A.-M., 2011, ``VLBI measurement of
secular aberration drift'', \aa 529, pp. A91--A97. 

Zharov, V., Sazhin, M., Sementsov, V., \& Sazhina, O. 2010,
``Long-term Variations of the EOP and ICRF2'', In: D. Behrend \&
K. D. Baver (Eds.), Proc. of the 2010 IVS 6th General Meeting,
pp. 290--294.

}

\end{document}